\documentclass[a4paper,amsmath,amssymb,prb,preprintnumbers,superscriptaddress,twocolumn,showpacs,longbibliography]{revtex4-1} 

\usepackage{graphicx}		

\usepackage{dcolumn}
\usepackage{bm}
\usepackage{amssymb}
\usepackage{epsfig}
\usepackage{gensymb}
\usepackage{multirow}
\usepackage{array}
\usepackage{ctable}

\usepackage{ulem}     

\newcommand{\lapprox}{{\scriptscriptstyle\stackrel{<}{\sim}}}

\newif\ifcom
\newif\ifdel

\comtrue               %
\deltrue                   %

\begin{document}

\title{Three-Axis Vector Nano Superconducting Quantum Interference Device}

\author{Mar\'ia Jos\'e Mart\'inez-P\'erez}
\email{mariajose.martinez@uni-tuebingen.de}
\affiliation{Physikalisches Institut -- Experimentalphysik II and Center for Quantum Science (CQ) in LISA${^+}$,
Universit\"at T\"ubingen, Auf der Morgenstelle 14, D-72076 T\"ubingen, Germany}

\author{Diego Gella}
\affiliation{Physikalisches Institut -- Experimentalphysik II and Center for Quantum Science (CQ) in LISA${^+}$,
Universit\"at T\"ubingen, Auf der Morgenstelle 14, D-72076 T\"ubingen, Germany}

\author{Benedikt M\"uller}
\affiliation{Physikalisches Institut -- Experimentalphysik II and Center for Quantum Science (CQ) in LISA${^+}$,
Universit\"at T\"ubingen, Auf der Morgenstelle 14, D-72076 T\"ubingen, Germany}

\author{Viacheslav Morosh}
\affiliation{Fachbereich 2.4 "Quantenelektronik", Physikalisch-Technische Bundesanstalt, Bundesallee 100, D-38116 Braunschweig, Germany}

\author{Roman W\"obling}
\affiliation{Physikalisches Institut -- Experimentalphysik II and Center for Quantum Science (CQ) in LISA${^+}$,
Universit\"at T\"ubingen, Auf der Morgenstelle 14, D-72076 T\"ubingen, Germany}

\author{Javier Ses\'e}
\affiliation{Laboratorio de Microscop\'ias Avanzadas (LMA), Instituto de Nanociencia de Arag\'on (INA), Universidad de Zaragoza, E-50018 Zaragoza, Spain}

\author{Oliver Kieler}
\affiliation{Fachbereich 2.4 "Quantenelektronik", Physikalisch-Technische Bundesanstalt, Bundesallee 100, D-38116 Braunschweig, Germany}

\author{Reinhold Kleiner}
\affiliation{Physikalisches Institut -- Experimentalphysik II and Center for Quantum Science (CQ) in LISA${^+}$,
Universit\"at T\"ubingen, Auf der Morgenstelle 14, D-72076 T\"ubingen, Germany}

\author{Dieter Koelle}
\affiliation{Physikalisches Institut -- Experimentalphysik II and Center for Quantum Science (CQ) in LISA${^+}$,
Universit\"at T\"ubingen, Auf der Morgenstelle 14, D-72076 T\"ubingen, Germany}

\date{\today}	

\begin{abstract}

We present the design, realization and performance of a three-axis vector nano Superconducting QUantum Interference Device (nanoSQUID).
It consists of three mutually orthogonal SQUID nanoloops that allow distinguishing the three components of the vector magnetic moment of individual nanoparticles placed at a specific position.
The device is based on Nb/HfTi/Nb Josephson junctions and exhibits linewidths of $\sim 250$\,nm and inner loop areas of $600 \times 90$ nm$^2$ and $500 \times 500$ nm$^2$.
Operation at temperature $T=4.2$\,K, under external magnetic fields perpendicular to the substrate plane up to $\sim 50$\,mT is demonstrated.
The experimental flux noise below $\sim 250\,{\rm n}\Phi_0/\sqrt{\rm Hz}$ in the white noise limit and the reduced dimensions lead to a total calculated spin sensitivity of $\sim 630\,\mu_{\rm B}/\sqrt{\rm Hz}$ and $\sim 70\,\mu_{\rm B}/\sqrt{\rm Hz}$ for the in-plane and out-of-plane components of the vector magnetic moment, respectively.
The potential of the device for studying three-dimensional properties of individual nanomagnets is discussed.

\end{abstract}



\maketitle


\section{Introduction}
\label{sec:Introduction}

Getting access to the magnetic properties of individual magnetic nanoparticles (MNPs) poses enormous technological challenges.
As a reward, one does not have to cope with troublesome inter-particle interactions or size-dependent dispersion effects, which facilitates enormously the interpretation of experimental results.
Moreover, single particle measurements give direct access to anisotropy properties of MNPs, which are hidden for measurements on ensembles of particles with randomly distributed orientation.\cite{MartinezPerez10,Silva05}

So far, different techniques have been developed and successfully applied to the investigation of individual MNPs or small local field sources in general.
Most of these approaches rely on sensing the local stray magnetic field created by the sample under study, by using \textit{e.g.}, micro- or nanoSQUIDs,\cite{Granata16,Schwarz15,Buchter15,Shibata15,Schmelz15,Vasyukov13,Arpaia14,Hazra14,Woelbing13,Levenson-Falk13,Buchter13,Bellido13,Granata13,Martinez-Perez11a,Hao11,Kirtley10,Giazotto10,Vohralik09,Huber08,Hao08,Wernsdorfer01,Jamet01,Wernsdorfer95,Awschalom92} micro-Hall magnetometers,\cite{Lipert10,Kent94} magnetic sensors based on NV-centers in diamond\cite{Thiel16,Schaefer-Nolte14,Rondin13} or magnetic force microscopes.\cite{Buchter15,Buchter13,Nagel13,Rugar04,Shinjo00}
Other probes, \textit{e.g.}, cantilever and torque magnetometers,\cite{Buchter15,Ganzhorn13,Nagel13,Buchter13,Stipe01} are sensitive to the Lorentz force exerted by the external magnetic field on the whole MNP.

For all magnetometers mentioned above, information on just one vector component of the magnetic moment $\bm\mu$ of a MNP can be extracted.
Yet, studies on the static and dynamic properties of individual MNPs would benefit enormously from the ability to distinguish simultaneously the three orthogonal components of $\bm\mu$.
This is so since real nanomagnets are three-dimensional objects, usually well described by an easy axis of the magnetization, but often exhibiting additional hard/intermediate axes or higher-order anisotropy terms.
Magnetization reversal of real MNPs also occurs in a three-dimensional space, as described by the classical theories of uniform (Stoner-Wohlfarth) \cite{Thiaville00,StonerWohlfarth} and non-uniform spin rotation.\cite{Aharoni66}
More complex dynamic mechanisms are also observed experimentally including the formation and evolution of topological magnetic states\cite{Buchter13} and the nucleation and propagation of reversed domains.\cite{Buchter15}

To date, few examples can be found in the literature in which three-axial detection of small magnetic signals has been achieved.
This was done by combining planar and vertical microHall-probes\cite{Schott00} or assembling together three single-axis SQUID microloops.\cite{Miyajima15,Ketchen97}
Further downsizing of these devices, which can significantly improve their sensitivity, is however still awaiting. This is mainly due to technical limitations in the fabrication of nanoscopic three-dimensional architectures.  

Very recently, an encouraging step towards this direction has been achieved by fabricating a double-loop nanoSQUID, patterned on the apex of a nanopipette\cite{Anahory14}. 
This device allowed to distinguish between the out-of-plane and in-plane components of the captured magnetic flux with $\sim 100$\,nm resolution, but only upon applying different external magnetic fields.

Here we present an ultra-sensitive three-axis vector nanoSQUID, fabricated on a planar substrate and operating at temperature $T=4.2$\,K.
The device is based on Nb/HfTi/Nb tri-layer Josephson junctions.\cite{Hagedorn06}
This technology involves electron beam lithography and chemical-mechanical polishing, which offers a very high degree of flexibility in realizing complex nanoSQUID layouts.
It allows the fabrication of planar gradiometers or stripline nanoSQUIDs, with sub-$100$\,nm resolution, in which the loop lies parallel or perpendicular to the substrate plane.\cite{Woelbing13,Nagel11a}
Thanks to this flexibility we have succeeded in fabricating three close-lying orthogonal nanoSQUID loops, allowing the simultaneous detection of the three vector components of $\bm\mu=(\mu_x,\mu_y,\mu_z)$ of a MNP placed at a specific position $\bm r_{\rm NP}$.
All three nanoSQUIDs operate independently and their voltage ($V$)-to-flux ($\Phi$) transfer function can be linearized by means of applying on-chip modulation currents $I_{\rm mod}$ for flux-locked loop (FLL) operation.\cite{Drung03}
Additionally, moderate magnetic fields up to  $\mu_0 H \sim 50$\,mT can be applied  perpendicular to the substrate plane, without degrading SQUID performance.
These nanoSQUIDs exhibit a measured flux noise below $250\,{\rm n}\Phi_0/\sqrt{\rm Hz}$ in the white noise regime (above a few 100\,Hz). 
The latter leads to spin sensitivities of  $\sim 610$, 650 and $70\,\mu_{\rm B}/\sqrt{\rm Hz}$ for the $\mu_x$, $\mu_y$ and $\mu_z$ components, respectively, of a MNP located at $\bm r_{\rm NP}=(0,0,0)$  ($\Phi_0$ is the magnetic flux quantum and $\mu_{\rm B}$ is the Bohr magneton).
As we demonstrate here, our device represents a valuable tool in the investigation of single MNPs providing information on, \textit{e.g.}, their three-dimensional anisotropy and the occurrence of coherent or non-uniform magnetic configurations.

\section{Results and discussion}
\label{sec:ResultsAndDiscussion }

\subsection{Sample fabrication and layout}
\label{subsec:SampleFabricationAndLayout}

\begin{figure}[t]
\includegraphics[width=5.5cm]{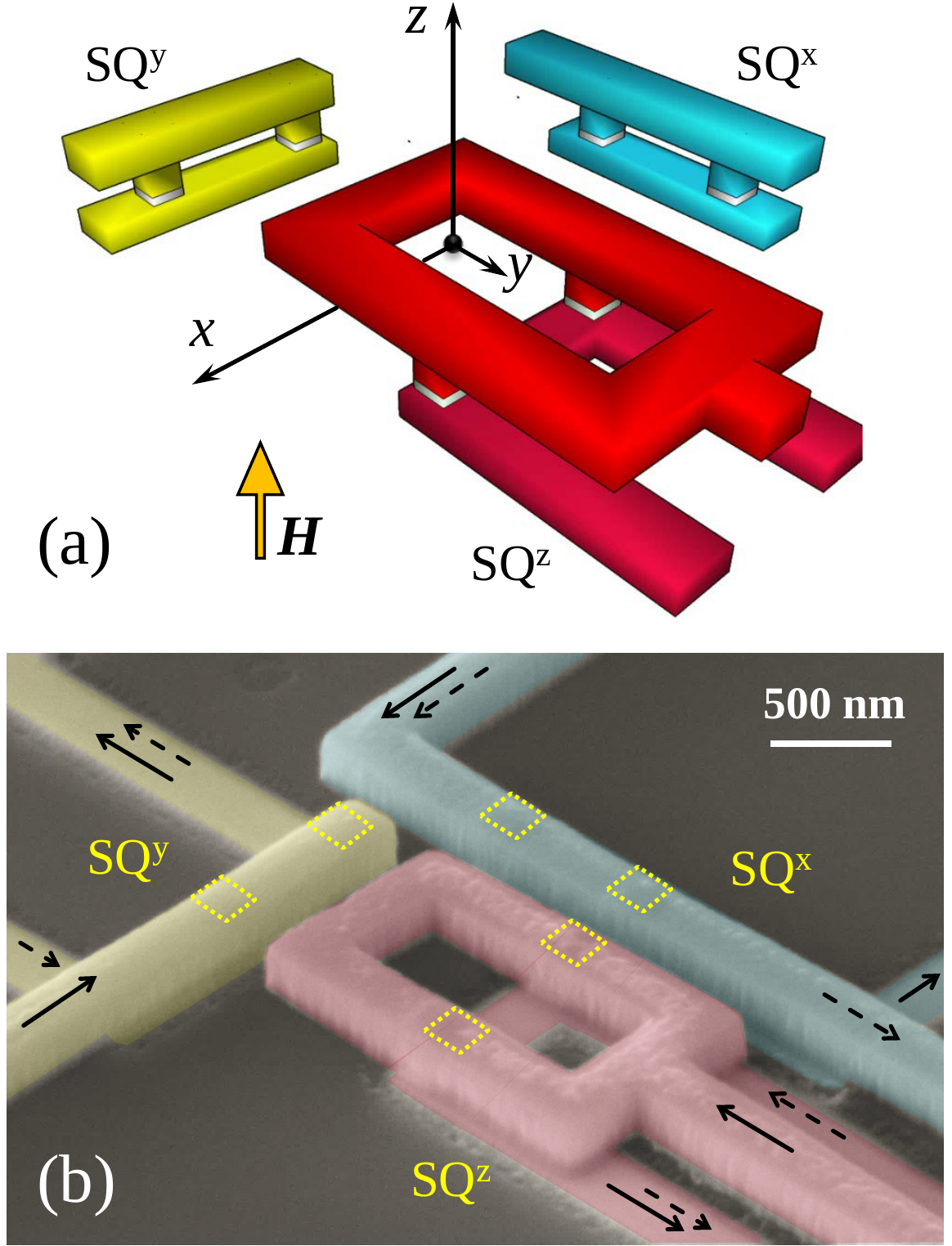}
\caption{(a) Schematic representation of the three-axis vector nanoSQUID consisting of three mutually orthogonal nanoloops.
SQ$^x$, SQ$^y$ and SQ$^z$ are used to detect the $\mu_x$, $\mu_y$ and $\mu_z$ components, respectively, of the magnetic moment $\bm\mu$ of an MNP.
The external magnetic field $\bm H$ is applied along $\hat{\bm e}_z$.
(b) False colored SEM image of a typical device.
Yellow dashed squares indicate the position of the Josephson junctions.
Black solid and dashed arrows indicate the direction of bias currents $I_{\rm b}$ and modulation currents $I_{\rm mod}$, respectively.
}
\label{Fig1}
\end{figure}

A scheme of the three-axis nanoSQUID is shown in Fig.~\ref{Fig1}(a).
Two perpendicular stripline nanoloops, SQ$^x$ and SQ$^y$, are devoted to measure the $x$ and $y$ components of $\bm\mu$, respectively.
The $z$ component of the magnetic moment is sensed by a third planar first-order gradiometer, SQ$^z$, designed to be insensitive to uniform magnetic fields applied along $\hat{\bm e}_z$ but sensitive to the imbalance produced by a small magnetic signal in one of the two SQUID loops.
Strictly speaking, the device reveals the three components of $\bm\mu$ if and only if the magnetic moment is placed at the intersection between the three nanoloop axes.
In practice, this position approaches $\bm r_{\rm NP}=(0,0,0)$ as indicated by a black dot in Fig.~\ref{Fig1}(a).
We note that $z=0$ corresponds to the interface of the upper Nb layer and the SiO$_2$ layer, which separates top and bottom Nb.
Later on we will demonstrate that this constraint is actually flexible enough to realize three-axis magnetic detection of MNPs with finite volume, even if these are not positioned with extreme accuracy.

Figure \ref{Fig1}(b) shows a false colored scanning electron microscopy (SEM) image of a typical device.
The junction barriers are made of normal metallic HfTi layers with thickness $d_{\rm HfTi}\approx 22$\,nm.
The bottom and top Nb layer are, respectively, $ 160$ and $200$\,nm-thick and are separated by a $90$\,nm-thick SiO$_2$ layer.
Nb wirings are $250$\,nm wide and the Josephson junctions are square-shaped with area $150 \times 150\,{\rm nm}^2$.
The inner loop area of SQ$^x$ and SQ$^y$ corresponds to $600 \times 90\,{\rm nm}^2$ whereas SQ$^z$ consist of  two parallel-connected loops with inner area of $500 \times 500\,{\rm nm}^2$.
This configuration allows the application of moderate homogeneous magnetic fields along $\hat{\bm e}_z$ that do not couple any flux neither to the nanoloops  of SQ$^x$ and SQ$^y$ nor to the junctions in the $(x-y)$-plane of all three nanoSQUIDs.

The bias currents $I_b$ and modulation currents $I_{\rm mod}$ flow as indicated in Fig.~\ref{Fig1}(b) by black solid and dashed arrows, respectively.
The latter are used to couple flux to each nanoSQUID individually, so to linearize their flux-to-voltage transfer function in FLL operation.

\subsection{Electric transport and noise data}
\label{subsec:ElectricTransportAndNoiseData}

\begin{figure}[t]
\includegraphics[width=\columnwidth]{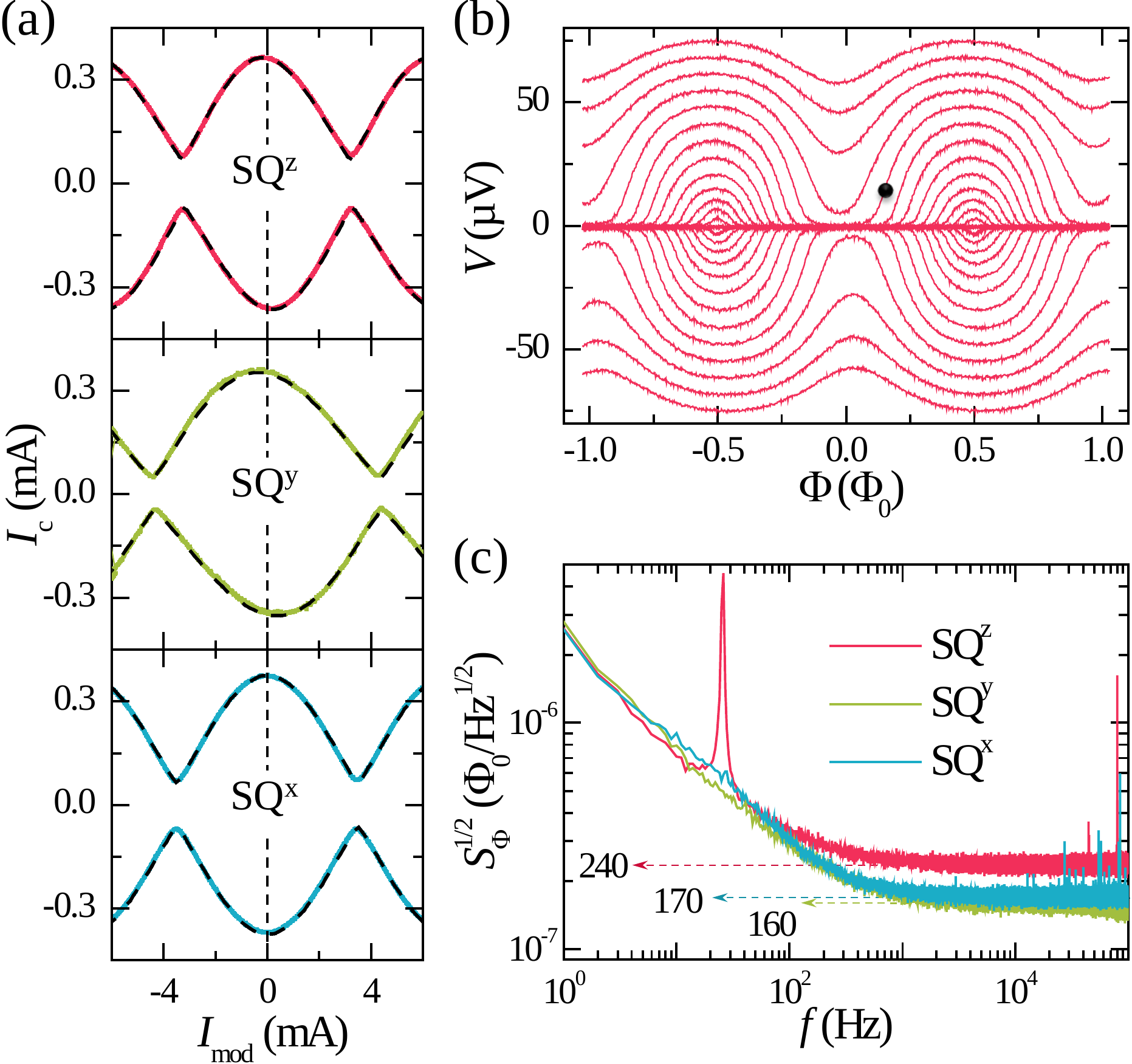}
\caption{Transport and noise characteristics of device A2.
(a) Measured (colored solid) and simulated (black dashed) modulation of the maximum critical current of the three nanoSQUIDs.
(b) $V (\Phi)$ measured for SQ$^z$ with $I_{\rm b}=-466 \ldots 471\,\mu$A (in $\sim 33.5\,\mu$A steps).
The black dot indicates the optimum working point with $V_{\Phi} \approx 330\,\mu{\rm V}/\Phi_0$ obtained for $I_{\rm b}=337\,\mu$A.
(c) Spectral density of rms flux noise measured for all three nanoSQUIDs in FLL-mode with an SSA.
Dashed arrows indicate the white noise values of $\sqrt{S_{\Phi}}$ in units of n$\Phi_0/\sqrt{\rm Hz}$.
}
\label{Fig2}
\end{figure}

The Nb/HfTi/Nb junctions have typical critical current densities $j_c \approx 550-850\,{\rm kA/cm}^2$ at $T=4.2$\,K and resistance times junction area $\rho_n \approx 9\,{\rm m}\Omega \mu{\rm m}^2$.
As a result, large characteristic voltages up to $V_c = j_c \rho_n \approx 60\,\mu$V can be obtained.
These junctions are intrinsically shunted providing, therefore, non-hysteretic current-voltage characteristics.\cite{Woelbing13,Nagel11a}

Electric transport data of a typical device are shown in Fig.~\ref{Fig2}.
From the period of the maximum critical current  $I_c(I_{\rm mod})$ shown in panel (a) we can deduce the mutual inductance $M^i\equiv \Phi^i/I_{\rm mod}^i$ between SQ$^i$ and its corresponding modulation line ($i=x,y,z$).
Asymmetries observed in these data for positive and negative bias current arise from the asymmetric distribution of $I_{\rm b}$ [see black solid arrows in Fig.~\ref{Fig1}(b)].
The strongest asymmetry is found  for SQ$^y$, which is attributed to the sharp corner in the bottom Nb strip right below one of the two Josephson junctions (see Fig.~\ref{Fig1}, upper right junction of SQ$^y$).
Numerically calculated curves based on the resistively and capacitively shunted junction (RCSJ)-model, including thermal noise, are fitted to these experimental data in order to estimate $\beta_L\equiv 2I_0L/\Phi_0$ and $I_c\equiv 2I_0$ [black dashed lines in Fig.~\ref{Fig2}(a)].
Here, $I_0$ is the average critical current of the two junctions intersecting the nanoloop, and $L$ is its inductance.
Asymmetric biasing is included in the model through an inductance asymmetry $\alpha_L\equiv (L_2-L_1)/(L_1+L_2)$ where $L_1$ and $L_2$ are the inductances of the two SQUID arms.
On the other hand, the maximum transfer coefficient $V_{\Phi} \equiv \partial V/\partial \Phi |_{\rm max}$ can be experimentally determined by coupling $\Phi$ via $I_{\rm mod}$ and measuring  the resulting $V(\Phi)$  for different $I_{\rm b}$ as shown in Fig.~\ref{Fig2}(b).
Following this approach, we have characterized a number of devices obtaining very low dispersion.
Few examples are provided in Table \ref{table1}, which gives evidence of the high quality and reproducibility of the fabrication process.

Finally, cross-talking between the three nanoSQUIDs can be quantified by the mutual inductances $M^{ij}=\Phi^i/I_{\rm mod}^j$ ($i\neq j$), i.e., the flux $\Phi^i$ coupled to SQ$^i$ by the modulation current $I_{\rm mod}^j$ in SQ$^j$.
%
%
%
If the three orthogonal SQUIDs are operated in FLL, a signal $\Phi^j$ detected by SQ$^j$ will be compensated by the feedback current $I_{\rm mod}^j=\Phi^j/M^j$.
This will also couple the (cross-talk) flux $\Phi^{ij}=\Phi^jM^{ij}/M^j$ to SQ$^i$.
As $M^{ij}$ is typically two orders of magnitude below $M^j$, this effect is negligible in most cases (see Methods section).
Moreover, it can be avoided by operating the devices in open-loop readout.

The operation of the sensor upon externally applied magnetic fields $\bm H = H \hat{\bm e}_z$ was investigated as well.
For this purpose, the output voltage response of all three nanoSQUIDs operating in FLL-mode was recorded upon sweeping $H$ for a number of devices.
Under optimum conditions, a negligible flux is coupled to SQ$^x$ and SQ$^y$ whereas, due to imperfect balancing, SQ$^z$ couples $\sim 5\,{\rm m}\Phi_0/$mT.
This imbalance results mainly from the asymmetric Nb wiring surrounding SQ$^z$ and the intrinsic errors associated to the fabrication.
All sensors are fully operative up to $\sim 50$\,mT, where abrupt changes in the response of the device are observed.
This behavior is attributed to the entrance of Abrikosov vortices in the Nb wires close to the nanoSQUIDs as it has been observed in similar devices.\cite{Nagel13,Woelbing13}
%

\begin{table}[t]
\centering
\caption{Parameters extracted from simulations based on the RCSJ-model and experimentally measured $1/M$$^i$ and $V_{\Phi}$ for three different devices (A2, D5 and C3).
}
\label{table1}
\begingroup
\setlength{\tabcolsep}{3.0pt} 
\renewcommand{\arraystretch}{1.2} 
\begin{tabular}{lr ccccccc}
				     	&	      		& $1/M^i$	  	& $I_0$ 		& $V_c$	   	& $\beta_L$	& $L$	& $\alpha_L$	& $V_{\Phi}$ \\ 
				     	&  			& (mA$/\Phi_0$)	& ($\mu$A)	& ($\mu$V)	&			&(pH)	&			& ($\mu$V/$\Phi_0$)   \\ \hline\hline
\multirow{3}{*}{A2}	& SQ$^x$	& 7.0		  	&187     		& 67			& 0.20		& 1.0	& 0			& 340   \\
					& SQ$^y$	& 8.8			&176	  	& 62			& 0.14		& 0.8	& 0.60	     	& 390   \\
				  	& SQ$^z$	& 6.5			&183		& 66			& 0.22		& 1.2  	& 0.25		& 330   \\ \hline
\multirow{3}{*}{D5}	& SQ$^x$	& 7.7		  	&136		& 57			& 0.14		& 1.1	& 0			& 250   \\
				  	& SQ$^y$ 	& 9.0			&136	  	& 59		 	& 0.12		& 0.9  	& 0.75	     	& 260   \\
				     	& SQ$^z$	& 5.7			&145	  	& 58			& 0.16		& 1.1	& 0.35	     	& 240   \\ \hline
\multirow{3}{*}{C3}	& SQ$^x$	& 8.0		  	&120	    	& 55			& 0.20		& 1.7	& 0			& 120   \\
				     	& SQ$^y$	& 9.1			&128		& 54			& 0.32		& 2.6	& 0.40		& 110   \\
				     	& SQ$^z$	& 5.8			&134		& 57			& 0.18		& 1.4  	& 0.28		& 170   \\ 
\end{tabular}
\endgroup
\end{table}

Fig.~\ref{Fig2}(c) shows the spectral density of rms flux noise $\sqrt{S_\Phi}$ obtained with each nanoSQUID operating in FLL mode after low-temperature amplification using a commercial SQUID series array amplifier (SSA).
The peak observed at $f = 26$\,Hz for SQ$^z$ is attributed to mechanical vibrations.
Ubiquitous $1/f$-noise dominates $\sqrt{S_\Phi}$ for $f \lapprox 100$\,Hz in all three spectra.
Remarkably low  values are obtained in the white region, yielding $\sqrt{S_\Phi} \approx 170$, 160 and 240\,n$\Phi_0/\sqrt{\rm Hz}$ for SQ$^x$, SQ$^y$ and SQ$^z$, respectively.

The flux noise can be translated into the spin sensitivity $\sqrt{S_\mu}  \equiv  \sqrt{S_\Phi}/\phi_\mu$, which is the figure of merit of nanoSQUID sensors.
Here, the coupling factor $\phi_\mu \equiv \Phi _\mu/\mu$ is the magnetic flux $\Phi_\mu$ per magnetic moment $\mu=|\bm\mu|$, which is coupled to the SQUID from a MNP with magnetic moment $\bm\mu=\mu\,\hat{\bm e}_\mu$ placed at position $\bm r$.
The coupling factor can be calculated as $\phi_\mu(\hat{\bm e}_\mu,\bm r) = \hat{\bm e}_\mu \cdot \bm b(\bm r) $,
where  $\bm b(\bm r) \equiv \bm B_J/J$ is the normalized magnetic field created at position $\bm r$ by a supercurrent $J$ circulating in the nanoloop.\cite{Woelbing14,Nagel11}
We note that $\phi_\mu$ depends on both the particle position $\bm r$ (relative to the nanoloop) and the orientation $\hat{\bm e}_\mu$ of its magnetic moment.
We simulate $ \bm b(\bm r)$ by solving the London equations for the specific geometry of each nanoSQUID (see Methods section).
For a particle at position $\bm r_{\rm NP}=(0,0,0)$ [see Fig. \ref{Fig1}(a)] we obtain for SQ$^i$ spin sensitivities $\sqrt{S_\mu^i}\sim 610$, 650 and $70\,\mu_{\rm B}/\sqrt{\rm Hz}$ for $i=x,y,z$, respectively.
The spin sensitivity for SQ$^z$ is much better than for SQ$^x$ and SQ$^y$, because $\bm r_{\rm NP}$  is much closer to SQ$^z$ than to SQ$^x$ and SQ$^y$.
%

\begin{figure}[t]
\includegraphics[width=\columnwidth]{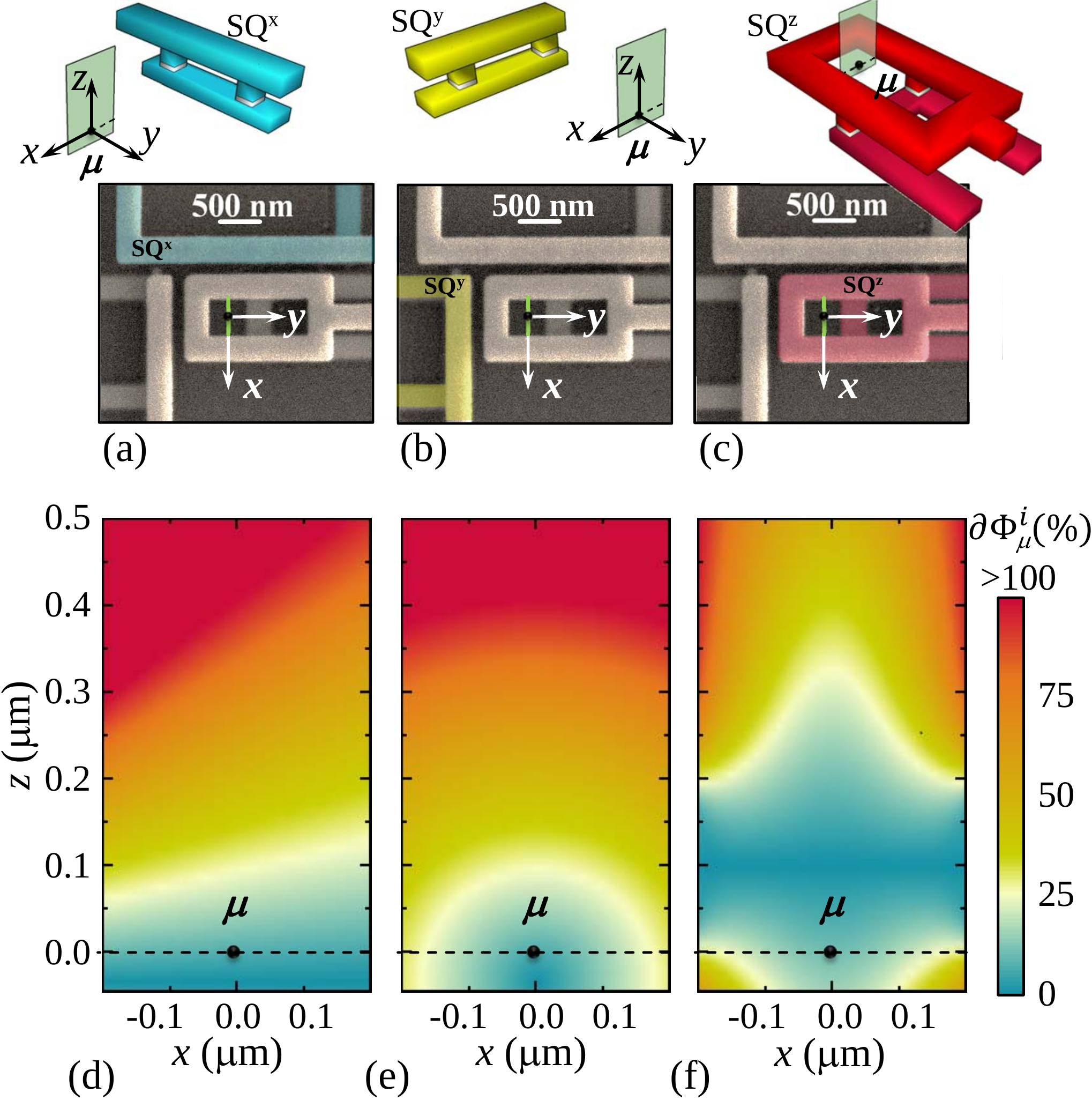}
\caption{(a)--(c): SEM images of the device with SQ$^x$ (a), SQ$^y$ (b) and SQ$^z$ (c) highlighted in false colors.
The green line indicates the $x$-$z$-plane at $y$=0 (shown schematically on top) for which the relative error $\partial  \Phi^i_\mu$ obtained for SQ$^x$, SQ$^y$ and SQ$^z$ is calculated in (d), (e) and (f), respectively.
The device works as a three-axis vector magnetometer when $\bm\mu$ is placed in regions with small $\partial  \Phi^i_\mu$.
Dashed lines correspond to $z=0$ (interface between SiO$_2$ and top Nb layer).
}
\label{Fig3}
\end{figure}

\begin{figure*}[t]
\includegraphics[width=14cm]{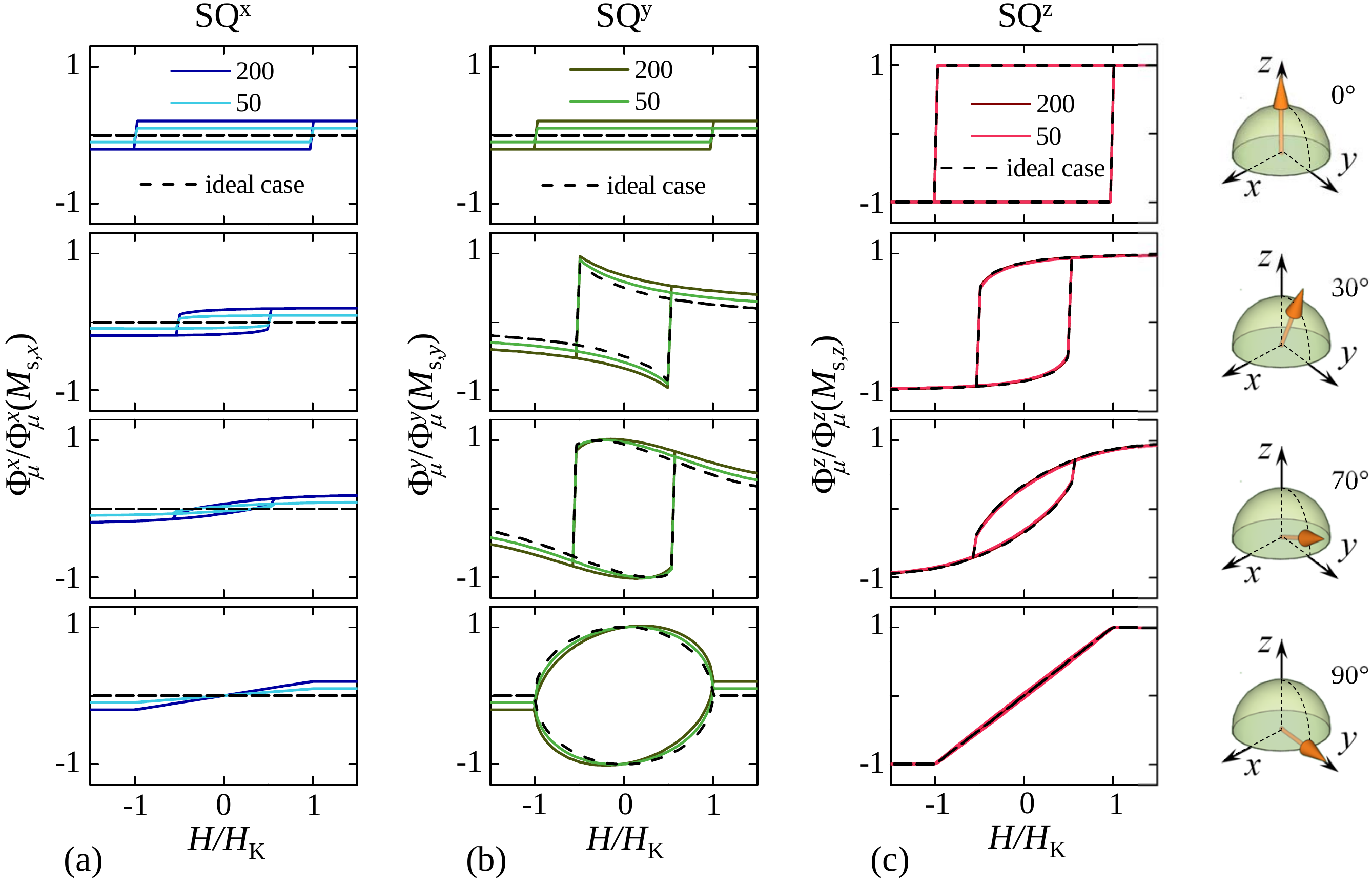}
\caption{Simulated magnetic hysteresis curves of a nanoparticle with magnetic moment $\bm\mu$ located at $\bm r_{\rm NP}=(0,0,0)$ as in Fig.~\ref{Fig3}.
The moment $\bm\mu$ couples magnetic flux $\Phi_\mu^x$, $\Phi_\mu^y$ and $\Phi_\mu^z$ to SQ$^x$ (a), SQ$^y$ (b) and SQ$^z$ (c), respectively.
$\bm H=H\hat{\bm e}_z$ with the particle's easy axis $\hat{\bm e}_{\rm K}$ lying at $0^\circ$, $30^\circ$, $70^\circ$ and $90^\circ$ (sketched in the right side of each panel).
$\Phi_\mu^i$ is normalized to the maximum possible flux in (a), (b) and (c) that is coupled when the particle is saturated along $\hat{\bm e}_x$ [$\Phi_\mu^x(M_{{\rm s}, x})$], $\hat{\bm e}_y$ [$\Phi_\mu^y(M_{{\rm s}, y})$] and $\hat{\bm e}_z$ [$\Phi_\mu^z(M_{{\rm s}, z})$], respectively ($M_{\rm s}$ is the saturation magnetization).
$H$ is normalized to the anisotropy field $H_{\rm K}$.
Black dashed lines correspond to an "ideal" case in which a point-like particle is coupled to an ideal three-axis magnetometer ($\partial  \Phi^i_\mu=0$) whereas colored solid lines correspond to a "realistic"' situation in which semi-spheres of radius $R=50$ and 200\,nm are measured with the device presented here.
MNPs are assumed to follow the Stoner-Wohlfarth model of magnetization reversal.
Different values of $R$ lead to a noticeably different behavior in (a) and (b), whereas all curves collapse into one in (c).
This stems from the fact that larger particles occupy regions with larger $\partial  \Phi^x_\mu$ and $\partial  \Phi^y_\mu$ as shown in Fig.~\ref{Fig3}(d) and (e).
}
\label{Fig4}
\end{figure*}

\subsection{Analysis of vector magnetometer performance}
\label{subsec:AnalysisOfVectorMagnetPerformance}

In the following we analyze the capability of this device to distinguish between the three components of $\bm\mu$.
For this purpose we write the normalized field $\bm b^i=(b_x^i,b_y^i,b_z^i)$ created by each SQUID SQ$^i$ as $\bm b^i = b_i^i\hat{\bm e}_i + b_{\perp i}^i\hat{\bm e}_{\perp i}$, \textit{i.e.}, we split this into a component along the $i$ direction and a component perpendicular to that, with $b_{\perp i}^i=\sqrt{(b_j^i)^2+(b_k^i)^2}$ ($i\neq j\neq k$). 
Ideally, for each of the three SQUIDs SQ$^i$, $b_i^i = |\bm b^i|\equiv b^i$, \textit{i.e.}, $b_{\perp i}^i=0$.
In that case, each SQUID SQ$^i$ is sensitive to the component $\mu_i$ only, and one can reconstruct the magnitude $\mu$ and orientation $\hat{\bm e}_\mu$ from the signals detected by the three orthogonal SQUIDs.

To quantify the deviation from that ideal case, we define the relative error flux $\partial  \Phi^i_\mu\equiv \Phi_{\mu,\perp i}^i / \Phi_{\mu,{\rm \parallel i}}^i$ made by nanoSQUID SQ$^i$.
Here, $\Phi_{\mu,{\rm \parallel i}}^i \equiv \mu \hat{\bm e}_i \cdot\bm b^i=\mu b_i^i$ relates to the ideal case in which the moment $\bm\mu$ is oriented along $\hat{\bm e}_i$.
In contrast, $\Phi_{\mu,\perp i}^i \equiv \mu\hat{\bm e}_{\perp i}\cdot\bm b^i = \mu b_{\perp i}^i$ corresponds to the worst case when the moment is oriented along $\hat{\bm e}_{\perp i}$, which yields the maximum error. 
Hence, the relative error flux is given by
$$\partial  \Phi^i_\mu\equiv\frac{\Phi_{\mu,\perp i}^i}{\Phi_{\mu,{\rm \parallel i}}^i} = \frac {\sqrt{(b_j^i)^2+(b_k^i)^2}}{b_i^i}\quad  ({\rm with}\; i\neq j\neq k) $$
This definition assures that $\partial  \Phi^i_\mu$ does not depend on the orientation $\hat{\bm e}_\mu$ of the magnetic moment of the particle, but only on its position $\bm r_{\rm NP}$.
The relative error flux for our device is first calculated at $\bm r_{\rm NP}=(0,0,0)$ giving $\partial \Phi^x_\mu = \partial \Phi^y_\mu \approx 7\,\%$ and $\partial \Phi^z _\mu\approx 4\,\%$.
Much better results can be obtained for SQ$^x$ and SQ$^y$ at $\bm r_{\rm NP}=(0,0,-0.035)\,\mu$m giving $\partial \Phi^x _\mu= \partial \Phi^y_\mu \approx 0.11\,\%$ and $\partial \Phi^z_\mu \approx 6\,\%$.
We note that this region becomes accessible after drilling a hole in the SiO$_2$ layer which is feasible by means of, \textit{e.g.}, focused ion beam milling.

We determine now deviations on the particle position that still lead to a tolerable level of error.
For this purpose $\partial  \Phi^i_\mu$ is calculated in the $x$-$z$-plane (at $y$=0) as indicated in Fig.~\ref{Fig3}(a), (b) and (c).
The results obtained for SQ$^x$, SQ$^y$ and SQ$^z$ are shown in (d), (e) and (f), respectively.
The white line in these color plots corresponds to $\partial  \Phi^i_\mu = 25\,\%$.
As it can be seen, SQ$^y$ imposes more severe restrictions on the particle position.
More specifically, $ \partial \Phi^y_\mu  \sim  10\,\%$ is obtained at $z=0$ and $x \approx \pm 55$\,nm, whereas $25\,\%$ results at $x \approx \pm 170$ nm.
Due to the symmetry of the problem the behavior of SQ$^x$ and SQ$^y$ is interchanged if one considers the $y$-$z$-plane.

We finish by showing how this device can indeed serve to provide full insight on the three-dimensional properties of MNPs of finite size and the mechanisms that lead to the magnetization reversal.
It will be instructive to start this discussion by focusing on the flux coupled by a point-like MNP to an ideal three-axis magnetometer, \textit{i.e.}, we assume $\partial  \Phi^i_\mu=0$ for $i=x,y,z$.
We consider for simplicity that the particle exhibits uniaxial anisotropy along a given direction $\hat{\bm e}_{\rm K}$, so that magnetic states pointing along $\pm \hat{\bm e}_{\rm K}$ are separated by an energy barrier.
In that case, the particle will exhibit a typical hysteretic behavior when sweeping the external magnetic field $\bm H = H \hat{\bm e}_z$.
This behavior will lead, however, to very different signals seen by each nanoSQUID, and those signals can strongly depend on the orientation of the easy axis with respect to the applied field direction.
This is represented in Fig.~\ref{Fig4}, where the flux $\Phi_\mu^i$ coupled to SQ$^i$ is plotted vs.~$H$ for $i=x$ (a), $y$ (b), $z$ (c) (dashed black lines).
The different panels correspond to different orientation of the easy axis, from $\hat{\bm e}_{\rm K} = \hat{\bm e}_z$ (top) to $\hat{\bm e}_{\rm K} = \hat{\bm e}_y$ (bottom), as sketched on the right side of Fig.~\ref{Fig4}.

Let us first consider the case in which the easy axis points along the externally applied magnetic field, i.e., $\hat{\bm e}_{\rm K} = \hat{\bm e}_z$.
As it can be seen, no flux is coupled to SQ$^x$ and SQ$^y$ as $\bm\mu$ lies always parallel to $ \hat{\bm e}_z$ whereas SQ$^z$ senses the maximum amount of flux possible. 
In the latter case, abrupt steps correspond to the switching of $\bm\mu$ between the $\pm \mu\hat{\bm e}_z$ states which leads to a typical squared-shaped hysteresis curve.
The situation changes dramatically  if one assumes that the easy axis points perpendicular to $\bm H = H \hat{\bm e}_z$. Under these circumstances, the particle's magnetic moment tilts progressively as the external magnetic field is swept so that no abrupt steps are observed in the hysteresis curves. This is exemplified in the bottom panels of Fig.~\ref{Fig4} which result when $\hat{\bm e}_{\rm K} = \hat{\bm e}_y$.
As it can be seen, $\Phi_\mu^x$ remains zero during the whole sweep whereas $\Phi_\mu^y=0$ is obtained only when the particle is saturated along $\hat{\bm e}_z$ leading to the maximum flux coupled by SQ$^z$. Remarkably, $\Phi_\mu^y$ reaches a maximum (minimum) at $H=0$ when $\bm\mu = +\mu \hat{\bm e}_y$ ($\bm\mu=-\mu \hat{\bm e}_y$) whereas $\Phi_\mu^z$ accounts for the progressive tilting of $\bm\mu$ as $H$ is swept.
Intermediate situations result when the easy axis points along different directions in space as it is exemplified in the middle panels.

Interestingly, a very similar behavior is observed when simulating a real experiment in which an extended MNP is measured using the three-axis nanoSQUID described here.
To illustrate this we have computed numerically $\Phi_\mu^i$ when semi-spheres with radius $R=50$ and 200\,nm centered at position $\bm r_{\rm NP}=(0,0,0)$ are considered (see Methods section).
As it can be seen in Fig.~\ref{Fig4} (solid lines) finite $\partial  \Phi^i_\mu\neq0$ and the particle's volume does not affect noticeably the flux coupled to SQ$^z$, whereas it slightly changes the flux coupled to SQ$^x$ and SQ$^y$.
This behavior can be easily understood, as the spatial extension of relatively large particles still remains in the region confined below the white line in Fig.~\ref{Fig3}(f), whereas they occupy zones with larger $\partial  \Phi_\mu^x$ and $\partial  \Phi_\mu^y$ in panels (d) and (e).
Still, our simulations demonstrate the operation of the device as a three-axis vector magnetometer even if relatively large MNPs are investigated.
The inspection of the hysteresis curves recorded simultaneously with all three nanoSQUIDs, together with the knowledge of the particle volume, allow extracting full information on the particle's anisotropy in a real experiment.

\subsection{Conclusions}
\label{subsec:Conclusions}

We have successfully fabricated three close-lying orthogonal nanoSQUIDs leading to the nanoscopic version of a three-axis vector magnetometer.
All three nanoSQUIDs can be operated simultaneously in open- or flux-locked loop mode to sense the stray magnetic field produced by an individual MNP located at position $\bm r_{\rm NP}$.
The device operates at $T=4.2$ K and is insensitive to the application of external magnetic fields perpendicular to the substrate plane (along $\hat{\bm e}_z$) up to $\sim 50$ mT. 
The latter can be used to induce the magnetization reversal of the MNP under study.
The limiting operation field can be increased in the future by improving the design. 
This implies reducing the linewidths so to increase the critical field for vortex entry and improving the balancing of the gradiometric nanoSQUID.

We have demonstrated the ability of this device to distinguish between the three orthogonal components of the vector magnetic moment by calculating the spatial dependence of the total relative error flux. 
The latter yields values below $10 \%$ for particles located at $\bm |r_{\rm NP}| \le 55$ nm. 
For $r_{\rm NP}=(0,0,0)$ we obtain a total spin sensitivity $\sim 610$, 650 and $70\,\mu_{\rm B}/\sqrt{\rm Hz}$ for the $x$, $y$ and $z$ components of $\bm\mu$, respectively. 
Finally, a model case has been described in which the three-axis vector nanoSQUID can be used to obtain full insight into the three-dimensional anisotropy of an extended MNP with diameter $\sim 100 - 400$ nm.
For this purpose, the signal captured by each nanoSQUID is used to reconstruct the magnitude and orientation of the magnetic moment during the magnetization reversal.

\section{Methods}
\label{sec:Methods}

\subsection{Sample Fabrication}
\label{sec:SampleFabrication}

The fabrication combines electron-beam lithography (EBL) and chemical-mechanical polishing (CMP).\cite{Hagedorn06}
A Si wafer with a $300$ nm-thick thermally oxidized layer is used as a substrate. 
An Al$_2$O$_3$ etch stop layer is first deposited by RF sputtering.
Then, the SNS tri-layer consisting of Nb/Hf$_{50\rm wt\%}$Ti$_{50\rm wt\%}$/Nb is sputtered \textit{in-situ}. 
The next step serves to define the SNS Josephson junctions by means of an Al etching mask defined by EBL and lift-off. 
The pattern is transferred to the Nb/HfTi/Nb tri-layer through reactive ion etching (RIE) in a SF$_6$ plasma and Ar ion beam acting on the counter Nb and HfTi layers, respectively. 
The bottom Nb layer is directly patterned using a negative EBL resist mask and SF$_6$-based RIE. 
In the following step, a 600 nm-thick layer of insulating SiO$_2$ is deposited through plasma enhanced chemical vapor deposition and subsequently polished through CMP. 
This process guarantees good wafer smoothing and electric contact to the Nb counter electrodes.
In the last step, the wiring Nb layer is sputtered and patterned using an EBL Al etching mask and SF$_6$-based RIE. 

\subsection{Measurement of electric transport properties and noise}
\label{subsec:MeasurementOfElectricTransportPropertiesAndNoise}

Current bias is performed by means of battery powered low-noise current sources and the output voltage is amplified at room temperature.
Each single nanoSQUID can be operated in flux-locked loop mode simultaneously, by using commercial three-channel SQUID readout electronics.
Additionally, the output signal can be amplified at low temperatures using commercial SQUID series arrays amplifiers.
High-field measurements are performed in a cryostat hosting a vector magnet, whereas noise measurements are performed in a magnetically and high-frequency shielded environment.
All  measurements described here were performed with the devices immersed in liquid $^4$He, at $T=4.2$\,K.

\subsection{Numerical simulations}
\label{subsec:NumericalSimulations}

Fitting of the $I_c(I_{\rm mod})$ experimental data is based on the RCSJ model.\cite{SQUIDhandbook}
The response of the SQUID is described by two coupled Langevin equations, $i/2+j=\beta_c \ddot{\delta}_{1}+ \dot{\delta_{1}} +\sin \delta_{1} +i_{\rm N1}$ and $i/2-j=\beta_c \ddot{\delta}_2 +\dot{\delta_2} +\sin \delta_2 +i_{\rm N2}$.
Here, $\delta_k(t)$ is the phase difference for the two junctions ($k=1,2$) and $i$ and $j$ are, respectively, the bias and circulating currents normalized to $I_0$.
Nyquist noise is included through two independent normalized current noise sources $i_{\rm N \textit{k}}$.
Additionally, $j \beta_L=(\delta_2-\delta_1)/\pi - 2 \varphi_{\rm ext} + \alpha_L \beta_L i/2$, where $ \varphi_{\rm ext}$ is the external flux normalized to $\Phi_0$.
Finally, $\beta_L\equiv 2I_0L/\Phi_0$, $\beta_c\equiv 2\pi I_0 R^2C/\Phi_0$, $\alpha_L\equiv (L_2-L_1)/(L_1+L_2)$, and $R$ and $C$ are the resistance and capacitance of the SQUID, respectively.
In the model, the total inductance of the loop $L=L_1+L_2$ accounts for both the geometrical and the kinetic contributions. 
The total dc voltage across the SQUID $V$ is calculated as the time average $V=\frac{1}{2} \left\langle U_1+U_2 \right\rangle$, where $U_k(t)=\frac{\Phi_0}{2 \pi}\dot{\delta}_k(t)$.
We emphasize here that the magnitude of $\beta_c$ does not affect the modulation of $I_c(I_{\rm mod})$ and, therefore, our estimation of $\beta_L$ and $I_c$. In all fittings $\beta_c = 0.5$ has been assumed for convenience, as in Chesca {\it et al.} \cite{SQUIDhandbook}.
%
%

\begin{table*}[t]
\centering
\caption{Calculated flux signals $\Phi_\mu^i$ and cross-talk signals $\Phi^{ij}$ for two different values of $R$ and parameters used for calculating cross-talk signals: average $\overline{M}^i$ (from values in Table \ref{table1}) and $M^{ij}$ measured for device D5.}
\label{table2}
\begingroup
\setlength{\tabcolsep}{5 pt} 
\renewcommand{\arraystretch}{1.2} 
\begin{tabular}{c|c|c|c|c|c|c|c|c|c|c|c|c}
\multicolumn{5}{c|}{}										& \multicolumn{4}{c|}{$R=50\,$nm}					& \multicolumn{4}{c}{$R=200\,$nm}\\								
		& $\overline{M}^i$	& \multicolumn{3}{c|}{$M^{ij}$}		& $\Phi_\mu^i$	& \multicolumn{3}{c|}{$\Phi^{ij}$}	& $\Phi_\mu^i$	& \multicolumn{3}{c}{$\Phi^{ij}$}\\
		& ($\Phi_0/$A)	& \multicolumn{3}{c|}{($\Phi_0/$A)}	& (m$\Phi_0$)	& \multicolumn{3}{c|}{(m$\Phi_0$)}	& (m$\Phi_0$)	& \multicolumn{3}{c}{(m$\Phi_0$)}\\\hline\hline
$j$		&				& $x$	&$y$	&$z$			& 				&$x$	&$y$	&$z$			&				& $x$	& $y$	& $z$\\\hline
SQ$^x$	& 132			& --		& 0.49	& 1.2			& 10				& --		& 0.044	& 1.4			& 600			& --  	& 2.6	& 72\\
SQ$^y$	& 112			& 1.5	& --		& 0.82			& 10				& 0.11	& -- 		& 0.98			& 600			& 6.8	& -- 		& 49\\
SQ$^z$	& 167			& 1.4	& 6.1	& --				& 200			& 0.11	& 0.55	& -- 				& $10^4$		& 6.4	& 33		& --
\end{tabular}
\endgroup
\end{table*}

For the estimation of the spin sensitivity and the relative error flux one needs to calculate the spatial distribution of $\textbf{B}^i_{J}$ created by each SQ$^i$.
For this purpose we have used the numerical simulation software 3D-MLSI\cite{Khapaev} which is based on a finite element method to solve the London equations in a superconductor with a given geometry, film thickness and London penetration depth ($\lambda_{\rm L}=90$ nm). 
$\bm b^x(\bm r)= \bm B^x_{J} /J$ and $ \bm b^y(\bm r)= \bm B^y_{J} /J$ with $J$ being the supercurrent in the nanoloop. 
For SQ$^z$ one needs to consider two circular currents $\pm J$ flowing around each nanoloop. 
The resulting normalized magnetic field is, in this case, $\bm b^z(\bm r) = \bm B^z_{J} /2 J$.

For the simulation of the hysteresis curves we consider first an ideal point-like MNP with magnetic moment $\bm\mu$ described by the polar coordinates $\hat{\bm e}_\mu=(1,\theta,\varphi)$ and characterized by one second-order anisotropy term. 
If both $\bm H$ and the easy axis lie in the $y$-$z$-plane the problem is reduced to the minimization of $e = \sin^2\phi-2h\cos(\phi + \Psi)$ in two dimensions ($\varphi=90^\circ$).
Here $e=E/U$ is the total energy normalized to the anisotropy barrier height, $h=H/H_K$ is the field normalized to the anisotropy field, $\Psi$ is the angle between $\bm H$ and the easy axis and $\phi=\theta-\Psi$ is the angle between $\bm\mu$ and the easy axis. 
Solutions of $\partial e/ \partial \phi =\partial e^2/ \partial^2 \phi=0$ for $\Psi=0^\circ$, $30^\circ$, $70^\circ$ and $90^\circ$ yield the values of
$$ \frac{ \Phi^i_\mu} {\Phi^i_\mu(M_{s,i})} = \frac {\hat{\bm e}_\mu \bm b^i (\bm r_{\rm NP})} {\hat{\bm e}_i \bm b^i (\bm r_{\rm NP})}$$
plotted in Fig.~\ref{Fig4}. 
Notice that, in this case, $\partial  \Phi^i_\mu=0$ so that $\bm b^i (\bm r_{\rm NP})=\hat{\bm e}_i b^i$ leading to $\Phi^x_\mu/\Phi^x_\mu(M_{s,x})=0$, $\Phi^y_\mu/\Phi^y_\mu(M_{s,y})=\sin \theta$ and $\Phi^z_\mu/\Phi^z_\mu(M_{s,z})=\cos \theta $.

For the simulation of extended particles we assume that all magnetic moments lie parallel to each other during the magnetization reversal. 
In this way, the exchange energy can be neglected and the expression for $e$ given above is still valid (Stoner-Wohlfarth model). 
Here, the second-order anisotropy term might also account for the shape anisotropy introduced by the magnetostatic energy. 
In this case one needs to integrate over the volume ($V_{\rm NP}$) of the whole MNP leading to
$$ \frac{ \Phi^i_\mu} {\Phi^i_\mu(M_{s,i})} = \frac {\int_{V_{\rm NP}} \hat{\bm e}_\mu \bm b^i (\bm r_{\rm}) dV} {\int_{V_{\rm NP}} \hat{\bm e}_i \bm b^i (\bm r_{\rm}) dV}.$$
Assuming, \textit{e.g.}, a semisphere made of hcp cobalt ($\mu=1.7$ $\mu_{\rm B}$/atom and density $8.9$ g/cm$^3$) one obtains $\Phi^x_\mu(M_{s,x})=\Phi^y_\mu(M_{s,y})\approx 10$ m$\Phi_0$ and $\Phi^z_\mu(M_{s,z})\approx 200$ m$\Phi_0$ for $R=50$ nm and $\Phi^x_\mu(M_{s,x})=\Phi^y_\mu(M_{s,y})\approx 0.6$ $\Phi_0$ and $\Phi^z_\mu(M_{s,z})\approx 10$ $\Phi_0$ for $R=200$ nm.
For these specific examples, we can compare the flux signals $\Phi_\mu^i$ quoted above with the corresponding cross-talk signals $\Phi^{ij}=\Phi_\mu^jM^{ij}/M^j$ appearing in FLL operation.
For the calculation of $\Phi^{ij}$, we used the experimentally determined values for $M^{ij}$ of device D5 and the average values $\overline{M}^i$ obtained from the measured values of all three devices  A2, D5 and C3 (from Table \ref{table1}).
These values are listed in Table \ref{table2} together with $\Phi_\mu^i$ and $\Phi^{ij}$.
We find that the cross-talk in FLL operation is on the percent level or even below, except for $\Phi^{xz}$ and $\Phi^{yz}$, where  it is around  10\,\%.

\acknowledgments 

M.~J.~M.-P.~acknowledges support by the Alexander von Humboldt Foundation, D.~G.~acknowledges support by the EU-FP6-COST Action MP1201.
This work is supported by the Nachwuchswissenschaftlerprogramm of the Universit\"at T\"ubingen, by the Deutsche Forschungsgemeinschaft (DFG) via Projects KO 1303/13-1, KI 698/3-1, and SFB/TRR 21 C2 and by the EU-FP6-COST Action MP1201.

\bibliography{VectorNanoSQUID}
\end{document}